\documentclass[intlimits,twoside,a4paper]{article}

\usepackage[cp1251]{inputenc}
\usepackage{multirow}
\usepackage{footnote}
\usepackage{tablefootnote}
\usepackage{booktabs}
\makesavenoteenv{tabular}
\usepackage{ragged2e}
\usepackage{soul}
\usepackage{color}
\usepackage[eqsecnum]{cmpj3}

\issue{2023}{26}{4}{43702}
\doinumber{10.5488/CMP.26.43702}
		
\title[First principles study of double perovskites $\text{Ca}_2\text{TMIrO}_6$ (TM=Fe, Co)]{Exploring the structural, electronic, magnetic, and magneto-optical properties of double perovskites $\text{Ca}_2\text{TMIrO}_6$ (TM=Fe, Co) through first principles study}
	
\author[I. Touaibia, A. Bouguerra, W. Guenez, F. Chemam]{I. Touaibia\refaddr{label1}, A. Bouguerra\refaddr{label2}, W. Guenez\refaddr{label1}, F. Chemam\orcid{0000-0003-0302-0608}\refaddr{label1}\thanks{Corresponding author:\email{fchemam@univ-tebessa.dz}} }
		\addresses{
		\addr{label1} Laboratory of Theoretical and Applied Physics (LPAT), Echahid Cheick Larbi Tebessi University, 12000 Tebessa, Algeria
		\addr{label2} Laboratory  of Physics of Matter and Radiation (LPMR), Department of Matter Sciences, University of Souk-Ahras, P. O. Box 1553, 41000 Souk-Ahras, 
		Algeria	}

	
\date{Received June 14, 2023, in final form September 12, 2023}

\begin{document}
\maketitle

\begin{abstract}
	This study is aimed at exploring the electronic, magnetic, and magneto-optical properties of double perovskites Ca$_2$FeIrO$_6$ and Ca$_2$CoIrO$_6$ {with monoclinic structure (space group P21/$c$) } in order to examine their potential applications in spintronic and photovoltaic devices. The calculations were done using the full-potential linearized augmented plane wave within the density functional theory. For the electronic exchange-correlation function, we used the generalized gradient approximation (GGA) and GGA+U (Hubbard potential), and spin-orbit coupling~(SOC). The study showed that Ca$_2$FeIrO$_6$ and Ca$_2$CoIrO$_6$ exhibit a monoclinic structure (space group P21/$c$). The structure relaxation shows an antiferromagnetic behavior in both systems with a magnetic moment of about 6.00$\mu_B$ for Ca$_2$FeIrO$_6$ and 4.00$\mu_B$ for Ca$_2$CoIrO$_6$ by using GGA+U approximation. The results of GGA and GGA+U predict the half-metallic behavior of Ca$_2$FeIrO$_6$ and Ca$_2$CoIrO$_6$. The magneto-optical polar Kerr effect (MOKE) was examined by studying the variation of Kerr and ellipticity rotation. The Kerr rotation angle is $1.3^{\circ}$  at 4.82~eV and $-1.21^{\circ}$ at 4.3~eV, and the ellipticity angle is $-1.21^{\circ}$ at 4.3~eV for Ca$_2$FeIrO$_6$. In the case of Ca$_2$CoIrO$_6$, the Kerr rotation angle is $-1.04^{\circ}$ at 4.05~eV; the significant Kerr rotation in both compounds may suggest the application of these materials in optoelectronics bias. The named compounds have a potential application in the field of spintronics and its devices, such as in optoelectronics technologies.
	\printkeywords
\end{abstract}

\section{Introduction:}
Since the fifties of the last century, double perovskite has been the subject of many studies; the interest in these materials is due to the important electronic and magnetic properties that range from metallic, insulator, semi-conductor, superconductor to half-metallic that were discovered in Sr$_2$FeMoO$_6$ compounds~\cite{Kobayasi, Pickett}. Double perovskites show significant magnetic properties. The coupling between electronic and magnetic properties, such as half metallicity with antiferromagnetic behavior, is of great interest, allowing them to be the most promising and critical multifunctional materials for use in the new generation of spintronics or optoelectronics and its devices~\cite{Wolf}. 
Double-perovskite oxides have the general formula A$_2$B$^\prime$B$^{\prime\prime}$O$_6$, where A is an alkaline-earth or rare-earth metal ion; B-site cations, {B$^\prime$ and B$^{\prime\prime}$},  are transition metals~\cite{Vasala}. The ideal structure of double perovskite is cubic, such as in Ba$_2$NiOsO$_6$~\cite{Vasala} Ba$_2$TMIrO$_6$ (TM$ = $Cr, Mn, and Fe)~\cite{Musa} and Ba$_2$BRuO$_6$ (B$= $Er, Tm)~\cite{Guenez}. However, the small-size radius of the cation A compared with the radius size of the other ions has an important effect on the structural properties of the material. When the A-site cation radius is smaller than the ideal, the structure compensates for the cation size mismatch by tilting and rotating the B$^\prime$O$_6$ and B$^{\prime\prime}$O$_6$ octahedral. This property is observed in many compounds with Ca, Sr, and La cations in the A-site and of monoclinic structure: Ca$_2$MnWO$_6$~\cite{Azed}, Sr$_2$YRuO$_6$~\cite{Bernardo}, Sr$_2$MReO$_6$ (M$= $Sc, Mn), Ca$_2$MReO$_6$ (M$= $Co, Cr, Fe, Mn)~\cite{Kato} and La$_2$BIrO$_6$ (B$ = $Mg, Mn, Co, Ni, Cu)~\cite{Currie}. 
Double perovskites can show both ferromagnetic (F) and antiferromagnetic (AF) behavior. This can be explained mainly by the super-exchange interaction between the neighboring transition metal ions B and B$^\prime$ via oxygen ions according to the Goodenough--Kanamori rule~\cite{Goodenough,Kanamori}. Xuedong et al.~\cite{Xuedong} found that the second nearest-neighboring~(2NN) Ir-Ir antiferromagnetic coupling is even stronger than the first nearest-neighboring~(1NN) Ni-Ir in double perovskites Sr$_2$NiIrO$_6$ and Sr$_2$ZnIrO$_6$. 
Ca$_2$FeIrO$_6$ was synthesized by Bufaial et al.~\cite{Bu}. Using experiments of magnetic susceptibility and X-ray powder diffraction, they found that the Ca$_2$FeIrO$_6$ compounds form in a monoclinic structure, space group P21/$n$, with the antiferromagnetic ground state. The theoretical study using the density functional theory was done by Bhandari et al.~\cite{Ram}, which predicts that the material should have a band gap of 0.13~eV and a significant effective magnetic moment of $6.68 \mu_B$ per unit cell.
Our work is aimed at studying the structural, electronic and magneto-optical  properties for the monoclinic structure of double perovskites Ca$_2$TMIrO$_6$ (TM=Fe, Co).

\section{Computational details} 
The full potential linear augmented wave method (FP-LAPW) within the first principles density functional theory (DFT) implemented in the WIEN2k program~\cite{Blaha} was applied to determine the structural, electronic, and magnetic properties of {Ca$_2$TMIrO$_6$ (TM=Fe, Co)}. The exchange-correlation potential was described using the PBE96 form of the generalized gradient approximation (GGA)~\cite{Perdew1,Perdew2}. The muffin-tin radii used were 2.07, 1.99, 1.99, 1.93, and 1.58 for Ca, Co, Fe, Ir, and O, respectively; we used 1000 $k$-points (230 $k$-points in the irreducible Brillouin zone), the value of separation energy is 7.180~Ry. The cut-off parameters are $RMT-k_{\rm{max}}=7$ and $RMT-G_{\rm{max}} = 12$. To explore the effect of spin-orbit coupling and on-site Coulomb repulsion, (DFT+SO) and (DFT+SO+U) were introduced. The added Hubbard potential $U$ was used with $U=1.5$~eV for Ir and $U=4$~eV for Co~\cite{Halder} and $U=5$~eV for Fe~\cite{Ram}. The convergence criterion for the self-consistent calculation was 0.0001~Ry for the total energies.

\begin{figure}[h]
	\centering
	\begin{minipage}{0.35\textwidth}
		\centering
		\includegraphics[width=\linewidth,height=7cm]{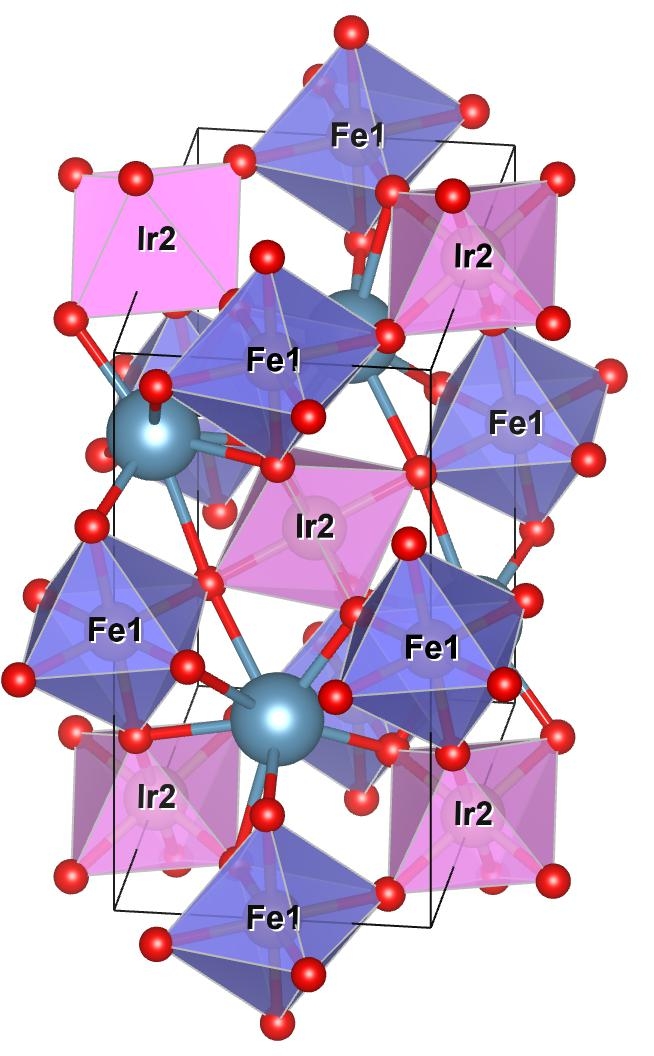}
	\end{minipage}
	\begin{minipage}{0.50\textwidth}
		\centering
		\includegraphics[width=\linewidth,height=7cm]{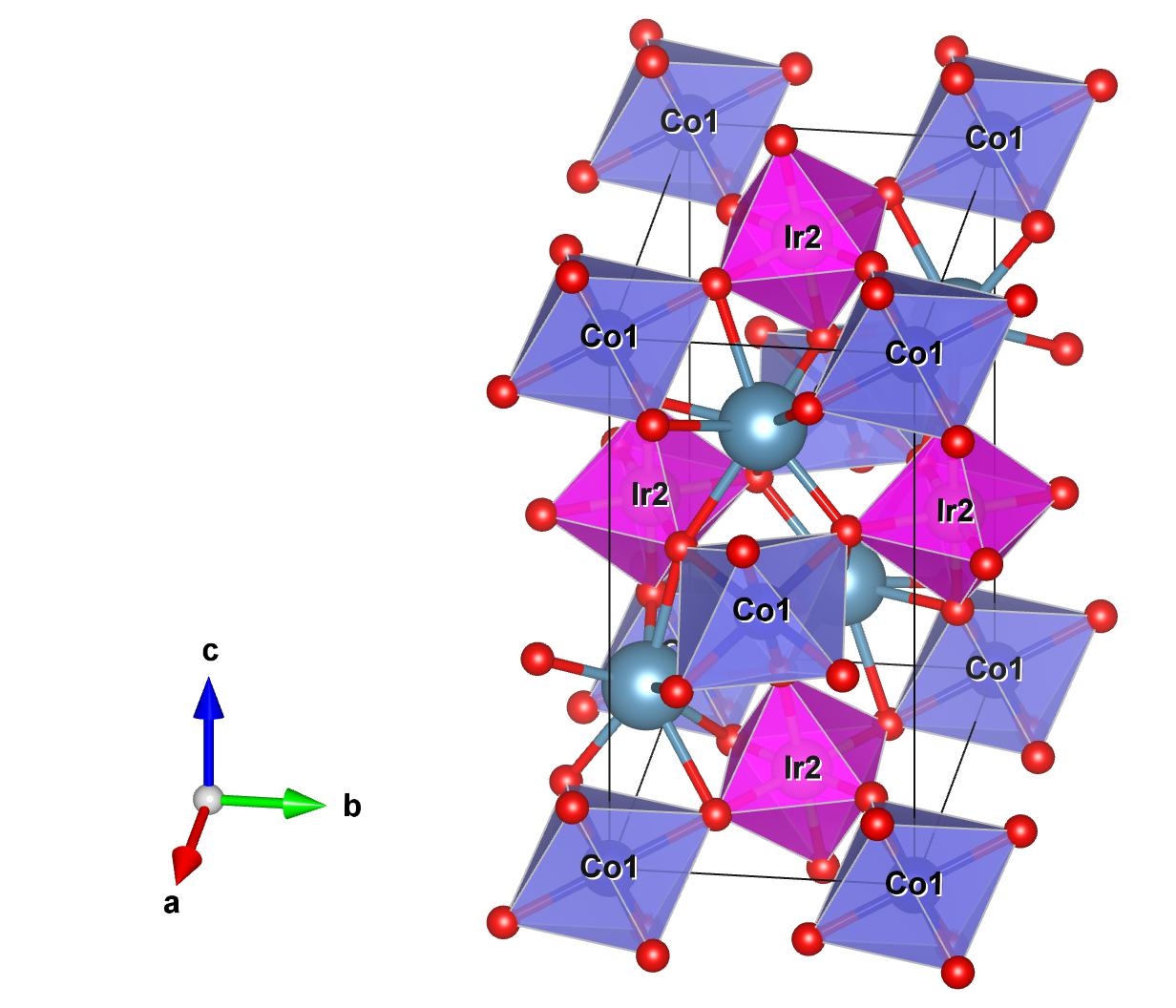}
	\end{minipage}
	\caption {(Colour online) Crystal structure of double perovskite $\text{Ca}_2\text{TMIrO}_6$ (TM=Fe, Co).}
	\label{fig:1}
\end{figure}

\section{Results and discussion}
\subsection{Structural properties}
To investigate  various physical properties of these compounds, we determine the stable ground states by studying the energy variation as a function of its unit cell volume. 
The results of the relaxation for lattice constants and volume for both compounds are summarized in table~\ref{tb1}, where we see that the ferrimagnetic (FIM) ground state is the most stable since the corresponding energy is minimum. As far as we know, this is the first study of these samples with a space group   (P21/$c$), whereas the recent other works~\cite{Bu, Ram, Halder} have explored similar samples characterized by a monoclinic structure with the space group (P21/$n$). 


\begin{table}[h!]
	\caption{Ferromagnetic (FM) and ferrimagnetic (FIM) equilibrium ordering with lattices constants.}
	\label{tb1}
	\begin{center}
	\footnotesize{
	\begin{tabular}{lllllll}
		\toprule
		& \multicolumn{2}{c}{Ca$_2$FeIrO$_6$}       &               & \multicolumn{2}{c}{Ca$_2$CoIrO$_6$}       
		&               \\
			\cmidrule(lr){2-4} \cmidrule(lr){5-7}
		& \multicolumn{1}{c}{FM} &                & FIM           & \multicolumn{1}{c}{FM} &                & FIM           \\
		\midrule
		$E$ (meV) $^{[1]}$      & $-1139804.2311$          &                & $-1139804.2435$ & $-1146368.9446$          &                & $-1146368.9457$ \\
		$\Delta E$ $^{[2]}$ &                        & 12.4           &               &                        & 1.1            &               \\
		$a$ (\AA)   &                        & 5.4156         &               &                        & 5.3772         &               \\
		$b$ (\AA)   &                        & 5.6045         &               &                        & 5.5995         &               \\
		$c$ (\AA)   &                        & 9.4344         &               &                        & 9.3577         &               \\
		Volume (\AA$^{3}$)  &                        & 1569.2624         &               &                        & 1544.5340         &               \\   
		$\beta$    &                        & 124.994$^{\circ}$ &               &                        & 124.632$^{\circ}$ &         
		\\ \hline 
		\multicolumn{7}{l}  {\footnotesize{$^{[1]}$ {Total energy}, $^{[2]}$ {$\Delta E=E_{\text{FM}}-E_{\text{FIM}}$}}.}
\end{tabular}
}
\end{center}
\end{table}


\subsection{Electronic properties}
\begin{figure}[h!]
	\centering
	\begin{minipage}{1.0\textwidth}
		\centering
		\includegraphics[width=\linewidth]{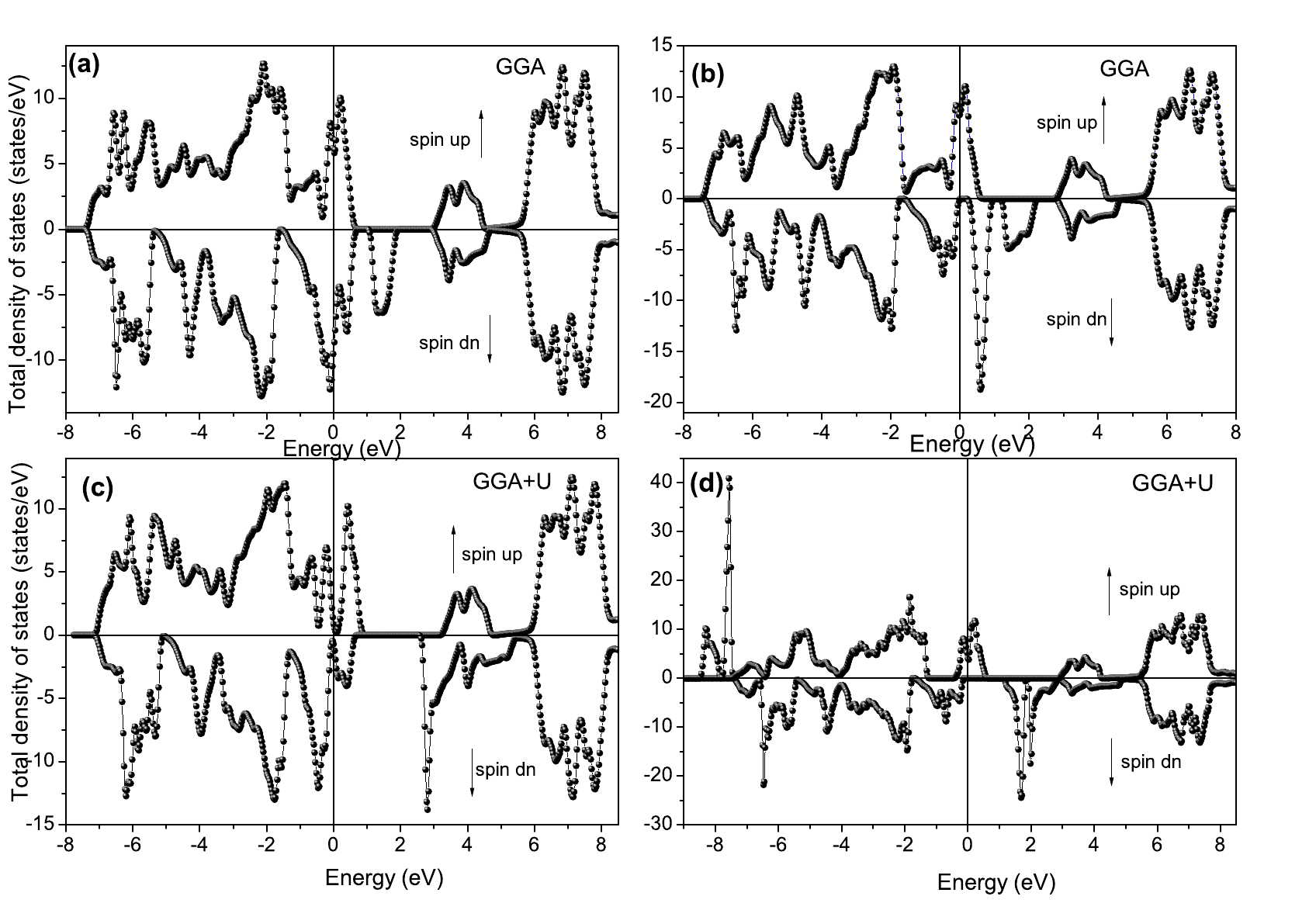}
	\end{minipage}%
	\caption{(Colour online) Total densities of states: $\text{Ca}_2\text{CoIrO}_6$ (a, c) using GGA and GGA+U,  $\text{Ca}_2\text{FeIrO}_6$ (b, d) using GGA and GGA+U. }
	\label{fig:2}
\end{figure}

\begin{figure}[h]
	\centering
	\begin{minipage}{1.0\textwidth}
		\centering
		\includegraphics[width=\linewidth]{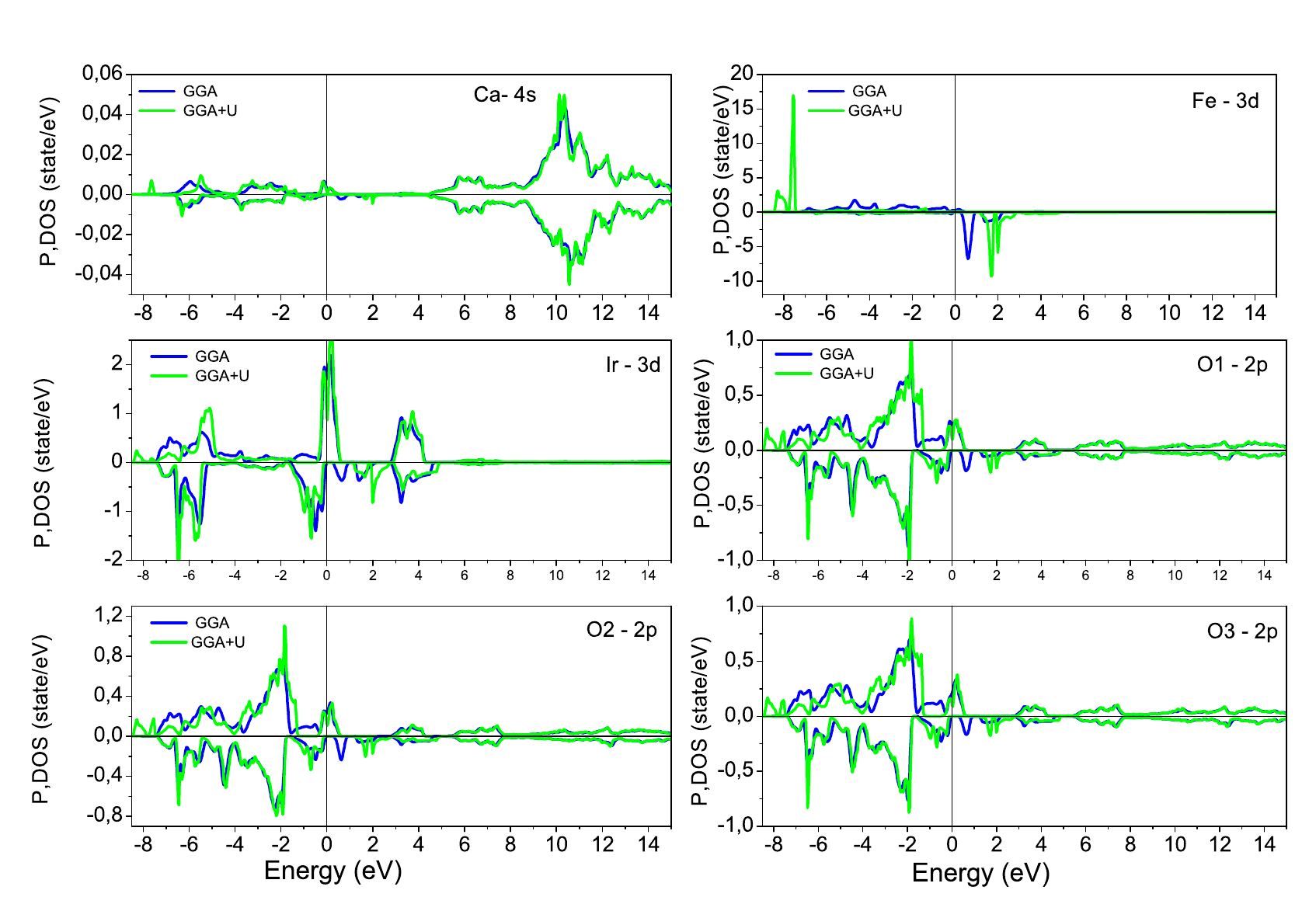}
	\end{minipage}%
	\caption{(Colour online) The projected density of states using GGA and GGA+U approximation at antiferromagnetic configuration utilizing ($U_{\rm{Fe}}=5$~eV, $U_{\rm{Ir}} = 1.5$~eV) of {Ca$_2$FeIrO$_6$.}   }
	\label{fig:3}
\end{figure}

\begin{figure}[h]
	\centering
	\begin{minipage}{1.0\textwidth}
		\centering
		\includegraphics[width=\linewidth]{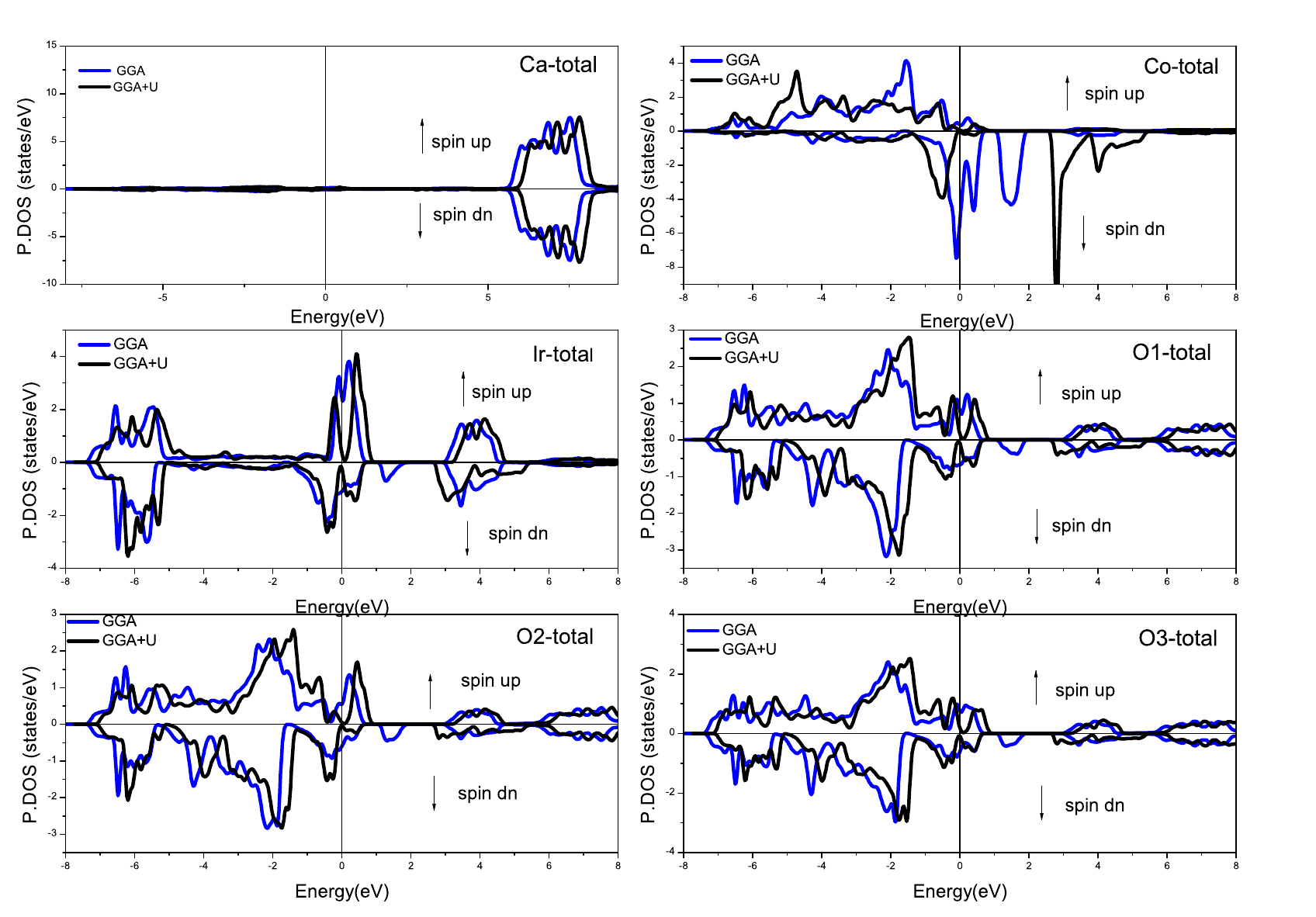}
	\end{minipage}%
	\caption{(Colour online) The projected density of states using GGA and GGA+U approximation at antiferromagnetic configuration utilizing ($U_{\rm{Co}}=5$~eV, $U_{\rm{Ir}} = 1.5$~eV ) of {Ca$_2$CoIrO$_6$.}  }
	\label{fig:4}
\end{figure}

Total density of state (DOS) in figure~\ref{fig:2} and partial density of state (PDOS) in figure~\ref{fig:3} and figure~\ref{fig:4} were computed in GGA and GGA+U approximations. 
{Figure~\ref{fig:2}(a,~c)} shows that Ca$_2$CoIrO$_6$ is a metal; however, in figure~\ref{fig:2}(b,~d), the Ca$_2$FeIrO$_6$ compound is half-metallic with $100\%$ polarization (metal in the spin-up and semiconductor in the spin-down).
The GGA and GGA+U approximation PDOS of Ca$_2$FeIrO$_6$ in figure~\ref{fig:3} shows the existance of different regions. The first region located in the deeper valence band between {$-7.47$ to 0~eV} is mainly due to the $d$ states of Ir with a small contribution of O ions. The second region is located around the Fermi level {0 to 2.55~eV}, and the majority of contributions are created by the hybridization of Fe($3d$)-O($2p$) and Ir($5d$)-O($2p$). In this region, it is evident that Fe~($3d$) and Ir ($5d$) ions are responsible for the half-metallic character around the Fermi level. The third region located at the bottom of the conduction band from 2.8~eV to 4.77~eV is mainly due to the $3d$ states of Ir and the $p$ states of oxygen. The upper part in the conducting band from 2.8~eV to the higher energies is due to the contributions from O~($2p$) and Ca~($4s$) states. 

From figure~\ref{fig:2}(a) and figure~\ref{fig:4}, the deep part of the valence band {$-7$ to 0~eV}  in Ca$_2$CoIrO$_6$ is due to the $5d$ states of Ir ions and $3d$ states of Co ions, which are dominant in this part, with a small contribution of oxygen ions. Around the Fermi level, between 0~eV and  0.95~eV, the majority of contributions are created by the hybridization of Co($3d$)-O($2p$) and Ir($5d$)-O($2p$). The Co~($3d$) and Ir~($5d$) are responsible for the half-metallic behavior around the Fermi level. In the bottom of the conduction band  {2.63 to  5.54~eV}, there is a hybridization between the $5d$ states of Ir and Co~($3d$) and the $p$ states of oxygen. The upper part in the conducting band {5.92 to 8.43~eV} comes from the contributions of O~($2p$) and Ca~($4s$) states. By inclusion of the on-site Coulomb interaction GGA+U, the band-gap in Ca$_2$FeIrO$_6$ is enlarged to 1.32~eV by shifting all states of unoccupied $d$ bands upward to higher energies and lowering the energy of occupied $d$ bands as seen in figure~\ref{fig:2}(d). However, the shape of the state density remains the same. The same observation is in the case of Ca$_2$CoIrO$_6$, where the inclusion of Hubbard-U potential leads to the change in the behavior of the material from metallic to half-metallic with a band gap of 0.107~eV in the spin-up direction. By contrast, in the spin-down direction, it presents a metallic character. This observation is similar to the other studies because the generalized gradient approximation underestimated the band gap~\cite{Ram,syed,das}.

\subsection{Magnetic properties}
The density of state calculations revealed that both compounds present a ferrimagnetic behavior. Since the magnetic moments of Fe and Co are aligned anti-parallel to that of Ir due to the hybridization between the Co and Fe~($3d$) states and Ir~($5d$) states on the one hand, and O~($2p$) on the other hand. Further, the values of the magnetic moments are not equal; this behavior can be explained via the super-exchange interaction through oxygen ions~\cite{Goodenough,Kanamori}. 

The magnetic moments of Fe and Co ions represent the main contributors to the total magnetic moment of the two compounds, which is estimated at 3.62$\mu_B$ and {2.44}$\mu_B$, respectively, as shown by the results of the calculations listed in table~\ref{tb2}.

The magnetic moment on oxygen atoms is generated from oxygen vacancies~\cite{Shu}, and Ca atoms carry a negligible magnetic moment.
The magnetic moments within GGA and GGA+U methods summarized in table~\ref{tb2} show that the magnetic moment of Ca$_2$FeIrO$_6$ is {6.02${\mu_B}$}. This value is relatively great due to the large spin-orbit and Coulomb interaction in ($3d$) and ($5d$) transition metals, whereas the magnetic moment of Ca$_2$CoIrO$_6$ is 3.72${\mu_B}$. These values are increased when we include the on-site Coulomb interaction with 4${\mu_B}$ for Ca$_2$CoIrO$_6$. The calculated total and partial magnetic moments on Ca$_2$FeIrO$_6$ and Ca$_2$CoIrO$_6$ for the magnetization along the $c$-axis are shown in table~\ref{tb2}.

\begingroup
\renewcommand*{\thefootnote}{\alph{footnote}}

\begin{table}[h!]
	\caption{The calculated total and partial magnetic moments (in units of Bohr's magneton, $\mu_B$)  with band gap spin up ($\uparrow$)/spin down ($\downarrow$) in~eV for $\text{Ca}_2\text{FeIrO}_6$ and $\text{Ca}_2\text{CoIrO}_6$.}
	\label{tb2}
		\begin{center}
		\begin{tabular}{lccclcccl}
		\toprule
		& \multicolumn{4}{c}{$\text{Ca}_2\text{FeIrO}_6$} & \multicolumn{4}{c}{$\text{Ca}_2\text{CoIrO}_6$} \\
		\cmidrule{2-5} \cmidrule{6-9}  
		& $M_{\text{Fe}}$ & $M_{\text{Ir}}$ & $M_{\text{tot}}$ & {gap} & $M_{\text{Co}}$ &  $M_{\text{Ir}}$& $M_{\text{tot}}$ & {gap} \\
		
		\midrule
		GGA & {3.62} & {$-0.66$} &{6.02} & $\uparrow${0}\, $\downarrow${0.383} & {2.44} & {$-0.66$} & {3.72} & $\uparrow${0}\, $\downarrow${0}\\[1ex]
		& {3.48\footnotemark[1]} & {$-0.55$\footnotemark[1]} & {6.00\footnotemark[1]} &  &  &  & & \\[1ex]
		GGA+U & {4.14}  & {$-0.88$} & {6.00}  & $\uparrow${0}\, $\downarrow${1.34}  & {2.62} &{$-0.42$}  & {4} & $\uparrow${0.107}\, $\downarrow${0} \\[1ex]
		& {4.12\footnotemark[1]}  & {$-0.87$\footnotemark[1]} & {5.99\footnotemark[1]}  &  & {2.743\footnotemark[2]} & {$-0.431$\footnotemark[2]}  & {2\footnotemark[2]} & $\uparrow${0.09\footnotemark[2]}\,$\downarrow${0} \\[1ex]
	\hline  		
		\multicolumn{9}{l}{	$^{(\rm a)}$\cite{Ram}, $^{(\rm b)}$\cite{Halder} }\\
	\end{tabular}
\end{center}
\end{table}

\endgroup
\subsection{Magneto-optical properties }
The studied double perovskite materials exhibit a large spin-orbit coupling and are characterized by important electronic properties such as half-metallic with ferrimagnetic behavior. The semi-conducting behavior has an essential role in the optical transfer properties; thus, it has an interesting MO effect. In industry, materials with huge MO effects are of great interest for use as devices in the fields such as optical telecommunications, reading the  heads in data, visualization, sensors, or other optoelectronic devices. Many compounds have been studied, such as Sr$_2$FeWO$_6$, which has the maximum polar Kerr rotation of $3.87^{\circ}$\cite{Shu},  and Ba$_2$B$^\prime$RuO$_6$ (B$^\prime$=Er, Tm) with a huge magneto-optical Kerr effect angle of $17.7^{\circ}$ and $5.6^{\circ}$ in the infrared region~\cite{Guenez}. Currently, no study presents the magneto-optical properties of the compounds included in this study. Therefore,  we first focused on the magneto-optical Kerr effect of these materials by performing theoretical calculations. The magnetization in polar Kerr effect is perpendicular
to the surface of the sample and parallel to $c$-axis.
The linear response of a system due to an external electromagnetic field with a small wave vector can be described with the complex dielectric function:
\begin{equation}
\begin{pmatrix}
\epsilon_{xx}&\epsilon_{xy}&$0$\\
-\epsilon_{xy}&\epsilon_{xx}&$0$\\
$0$&$0$&\epsilon_{zz}\\
\end{pmatrix},
\label{eqn:TensorE}
\end{equation}
where $\epsilon_{xx}$ and $\epsilon_{zz}$ are the diagonal components and $\epsilon_{xy}$ is the off-diagonal components of the dielectric tensor.
The optical conductivity tensor is related to the dielectric tensor by the equation:
\[ \epsilon_{\alpha\beta}(\omega)=\delta_{\alpha\beta}+\frac{4\piup \ri}{\omega}\sigma_{\alpha\beta}(\omega), \]
where:
\begin{equation} 
	\sigma_{\alpha\beta} (\omega)=\frac{-\ri e^2}{m^2\hbar V}\sum_k  \sum_{mm'}\frac{f(E_{mk})-f(E_{m'k})}{\omega_{mm'}}\frac{\Pi^\alpha_{m'm}\Pi^\beta_{mm'}}{(\omega-\omega_{mm'}+\ri\gamma)},
	\label{eq3}
\end{equation}

\noindent where,  $\Pi^\alpha_{m'm}$ and $\Pi^\beta_{mm'}$ are the optical dipolar matrix elements, $f(E_{mk})$ is the Fermi function, $E_{m'k}$ and $E_{mk}$ are Kohn-Sham energies at the position $k$ for two bands $ m', m$ respectively, so that $\hbar\omega_{mm'}=E_{mk}-E_{m'k}$, and $\gamma$ is the lifetime  parameter~\cite{Mahdi}.
The magneto-optical absorption is described by the imaginary part of the non-diagonal elements of the conductivity tensor.
The Kerr rotation $\theta_{k}$ and Kerr ellipticity $\epsilon_{k}$ were related to the optical conductivity by the relation given as~\cite{Bae,Ricinschi}:
\[ \theta_{k} + \ri\epsilon_{k}=\frac{-\sigma_{xy}}{\sigma_{xx}\sqrt{1+({2\piup \ri}/{\omega})\sigma_{xx}}}. \]

\begin{figure}[ht]
	\centering
	\begin{minipage}{0.48\textwidth}
		\centering
		\includegraphics[width=\linewidth,height=7cm]{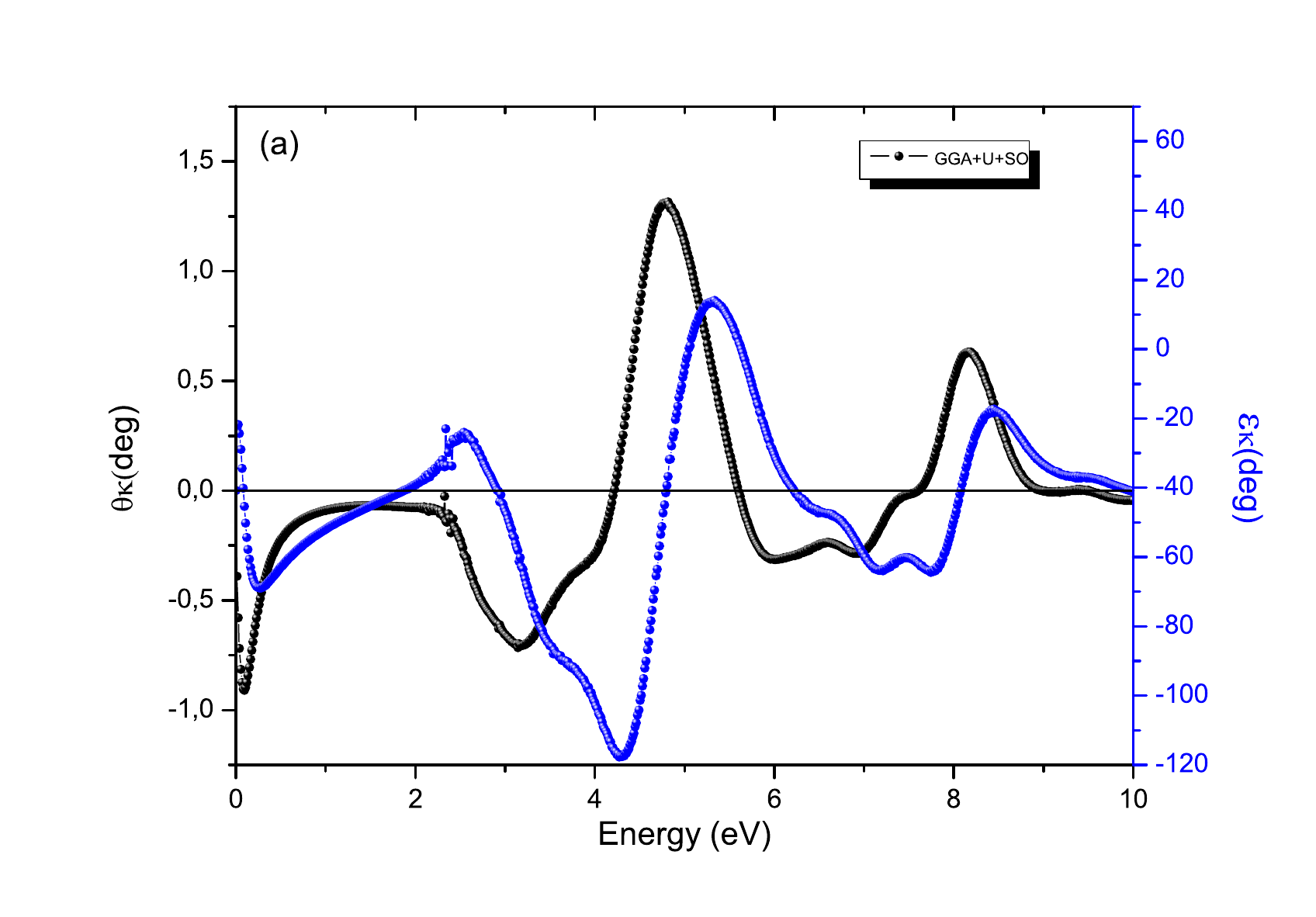}
	\end{minipage}
	\begin{minipage}{0.48\textwidth}
		\centering
		\includegraphics[width=\linewidth,height=7cm]{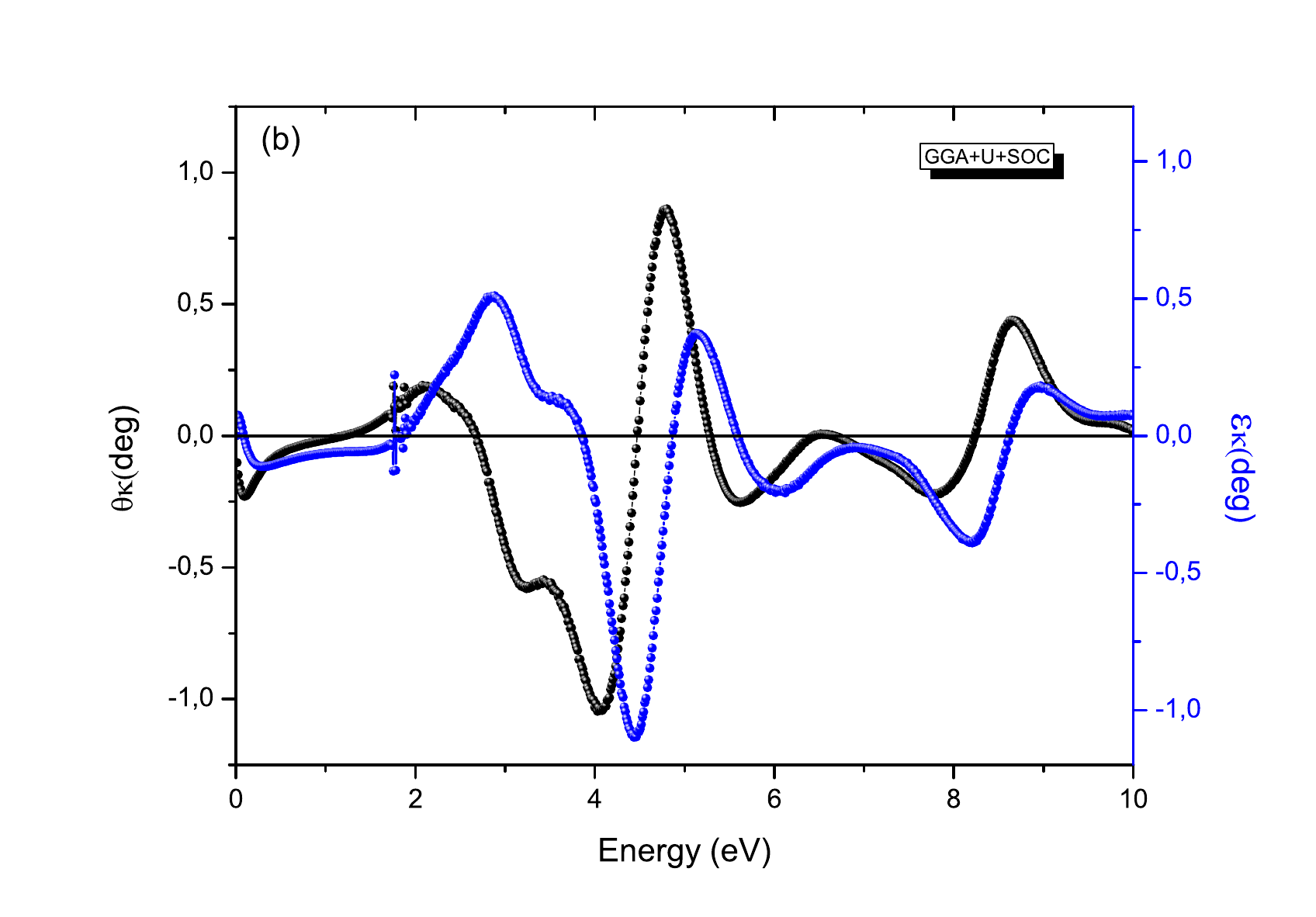}
	\end{minipage}
	\caption {(Colour online) The calculated polar Kerr angle $\theta_{k}$ and Kerr ellipticity $\epsilon_{k}$ as a function of photon energy (eV): (a) for Ca$_2$FeIrO$_6$ and (b) for Ca$_2$CoIrO$_6$.}
	\label{fig:5}
\end{figure}

Figure~\ref{fig:5} presents the Kerr rotation and ellipticity. We can see a peak of Kerr $\theta_{k}=-0.97{^\circ}$ at 0.1~eV and ellipticity rotation $\epsilon_{k}=-0.45{^\circ}$ at 0.28~eV with RCP dominance of polarized light for Ca$_{2}$FeIrO$_{6}$, while for Ca$_{2}$CoIrO$_{6}$, Kerr rotation $\theta_{k}=-0.23{^\circ}$   at 0.11~eV and $\epsilon_{k}= -0.11{^\circ}$ at 0.31~eV with RCP dominant. In this region of low energy, the Kerr rotation peaks are induced by the plasma resonance due to the metallic behavior of the compounds, as mentioned in the study of Feil and Haas~\cite{Feil}. We have the main peak of $1.3{^\circ}$ located at an incident photon energy of 4.82~eV (with large width of 1.38~eV) in Kerr rotation, and the main peak of $-1.21{^\circ}$ at 4.3~eV in ellipticity rotation for Ca$_{2}$FeIrO$_{6}$. For Ca$_{2}$CoIrO$_{6}$  the main peaks are $-1.1{^\circ}$ at 4.46~eV (ellipticity rotation) and $-1.04{^\circ}$ at 4.05~eV (Kerr rotation). Because of the Kramers-Kronig relation, we observe an alternation in the sign of $\theta_{k}$ and $\epsilon_{k}$. The calculated Kerr rotation was compared with other compounds as Ca$_2$FeWO$_6$ that have $\theta_{k}= 1.32{^\circ}$ at 1.89~eV and Ca$_2$FeReO$_6$ with $\theta_{k}= 1.04{^\circ}$ at 1.43~eV~\cite{Vidya}.
These large values of Kerr rotation in the range of 4.5 to 5~eV may suggest the potential application of these materials in ultraviolet laser light magneto-optical effect devices.

\begin{figure}[h!]
	\centering
	\begin{minipage}{1.0\textwidth}
		\centering
		\includegraphics[width=\linewidth]{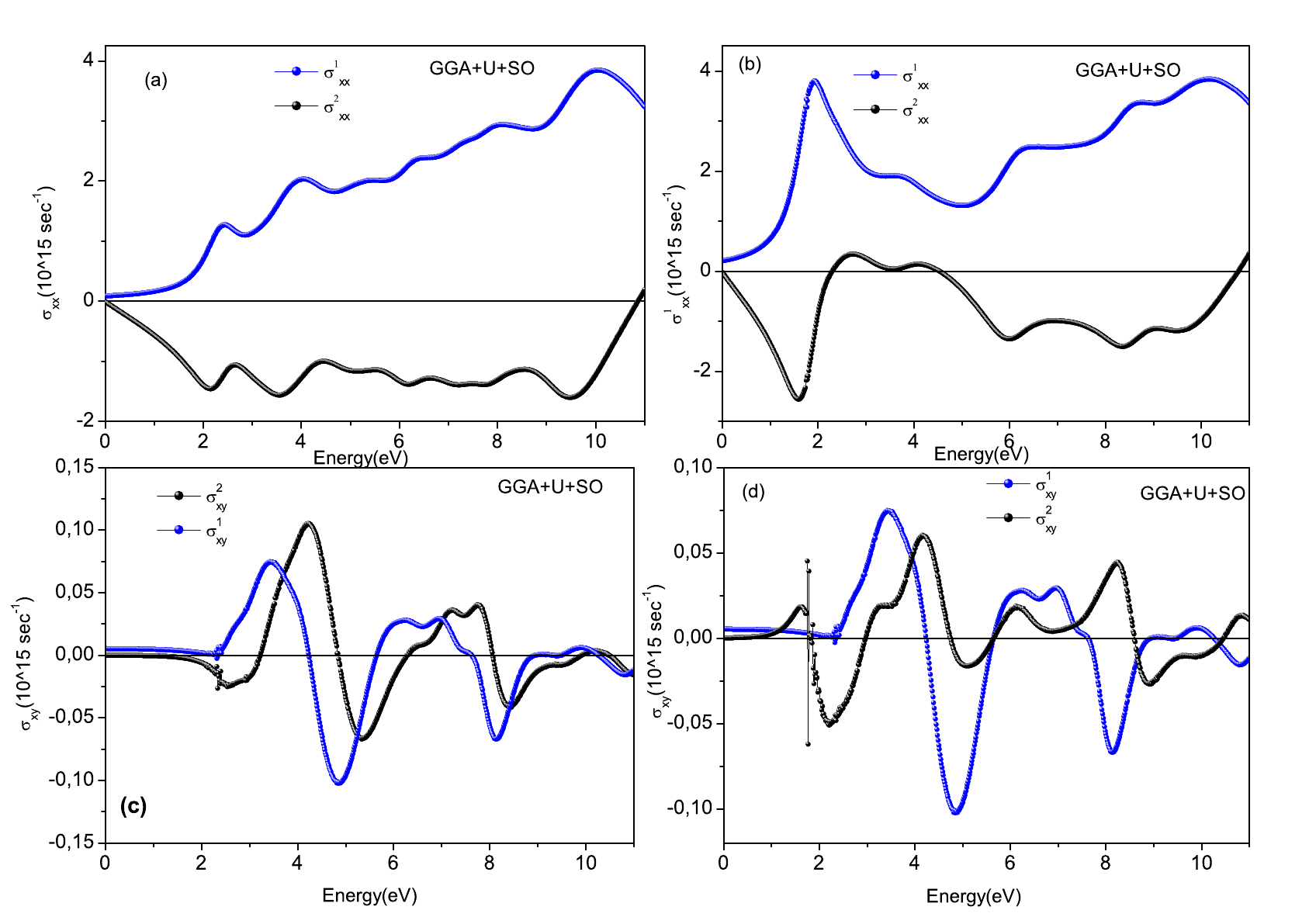}
	\end{minipage}%
	\caption{(Colour online) (a) and (b) represent the absorption $\sigma_{xx}^{1}$ and dispersive $\sigma_{xx}^{2}$ of optical conductivity, (c) and (d) off-diagonal optical conductivity $\sigma_{xy}^{1}$ and $\sigma_{xy}^{2}$ as a function of photon energy (eV) Ca$_2$FeIrO$_6$ and Ca$_2$CoIrO$_6$, respectively. }
	\label{fig:6}
\end{figure}
In the region of energy higher than 3.0~eV, the Kerr and ellipticity rotation peaks originate from the quite significant off-diagonal optical conductivity element.
Figure~\ref{fig:6} presents the calculated diagonal (the real part $\sigma_{xx}^{1}$ and the imaginary part $\sigma_{xx}^{2}$) and the off-diagonal (the real part $\sigma_{xy}^{1}$ and imaginary part~$\sigma_{xy}^{2}$) optical conductivity for both compounds in the range from 0 to 10~eV. In figure~\ref{fig:6}(a), we plotted the absorption spectra (the real part) and dispersive of the optical conductivity. We note that the band gap starts around 1.8~eV. In the region higher than 3.0~eV, the peaks in the two spectra are at the same positions; they originate from the $p$-$d$ inter-band transition~\cite{Jeng}.
The off-diagonal optical conductivity $\sigma_{xy}^{1}$ and $\sigma_{xy}^{2}$ are related to Kerr and ellipticity rotation. The main peak in the imaginary part of the off-diagonal optical conductivity is due to the principal inter-band transitions at all points.

\section{Conclusion}
The structural, electronic, magnetic, and magneto-optical properties of double perovskites Ca$_2$CoIrO$_6$ and Ca$_2$FeIrO$_6$ {with monoclinic structure (space group P21/$c$) } have been investigated using the first-principles DFT within the GGA and GGA+U approximation. We showed that 
{the configuration} with ferrimagnetic coupling is the most stable. GGA and GGA+U computations predict the half-metallic behavior for Ca$_2$FeIrO$_6$ and Ca$_2$CoIrO$_6$. The band gap was enlarged for Ca$_{2}$FeIrO$_6$ from 0.383~eV to 1.34~eV by including Hubbard-U electron correlation. In the case of Ca$_2$CoIrO$_6$,  the band gap is opened and presents the value of about 0.107~eV. Both compounds exhibit important electronic and magnetic properties that may suggest potential applications in spintronic devices. We predict a huge Kerr and ellipticity rotation $\theta_{k}=1.3{^\circ}$ at 4.82~eV for Ca$_2$FeIrO$_6$ and $-1.1{^\circ}$ at 4.46~eV for Ca$_2$CoIrO$_6$, which made them potential candidates for application in ultra-violet optoelectronic devices and hence promising multifunctional materials.

\section*{Acknowledgements}
We are very grateful to the Algerian Ministry of Higher Education and Scientific Research and
the DGRSDT for their financial support.



\ukrainianpart

\title{Дослідження структурних, електронних, магнітних і магнітооптичних властивостей  подвійних перовскітів  $\text{Ca}_2\text{TMIrO}_6$ (TM=Fe, Co) за першими принципами}

\author{І. Тоабія\refaddr{label1}, А. Боугерра\refaddr{label2}, В. Гуенез \refaddr{label1}, Ф. Чемам\refaddr{label1} }

\addresses{
\addr{label1} Лабораторія теоретичної та прикладної фізики (LPAT), Університет Ечахід Чеік Ларбі, 12000 Тебесса, Алжир
\addr{label2} Лабораторія фізики речовини та випромінювання (LPMR), Факультет матеріалознавства, Університет Сук-Аграс, BP 1553, 41000 Сук-Аграс, Алжир	}


%
%
%

\makeukrtitle

\begin{abstract}
	\tolerance=3000%
		Вивчаються електронні, магнітні та магнітооптичні властивості подвійних перовскітів Ca$_2$FeIrO$_6$ і Ca$_2$CoIrO$_6$ {моноклинної структури (просторова група P21/$c$) } з метою оцінки можливості їх потенційного застосування у спінтронних та фотоелектричних пристроях. Обчислення проведено з використанням повнопотеціального методу лінеаризованих приєднаних плоских хвиль в рамках теорії функціоналу гус\-тини. Для електронної обмінно-кореляційної функції використано узагальнені градієнтні наближення (GGA) та GGA+U (з потенціалом Габбарда), а також спін-орбітальний зв’язок~(SOC). Виявлено, що Ca$_2$FeIrO$_6$ і Ca$_2$CoIrO$_6$ мають моноклинну структуру  (просторова група P21/$c$). Релаксація структури в наближенні  GGA+U виявила антиферомагнітну поведінку в обидвох системах з магнітними моментами близько 6.00$\mu_B$ для Ca$_2$FeIrO$_6$ та 4.00$\mu_B$ для Ca$_2$CoIrO$_6$. Результати наближень GGA і GGA+U передбачають напівметалічну поведінку Ca$_2$FeIrO$_6$ та Ca$_2$CoIrO$_6$. Магнітооптичний полярний ефект Керра (MOKE) досліджувався шляхом вивчення змін в обертаннях поляризацій Керра та поляризації еліптичного типу.
	Для Ca$_2$FeIrO$_6$ кут повороту Керра складає  $1.3^{\circ}$  при 4.82~еВ і $-1.21^{\circ}$ при 4.3~еВ, а кут еліптичності $-1.21^{\circ}$ при 4.3~еВ. Для Ca$_2$CoIrO$_6$ кут повороту Керра становить $-1.04^{\circ}$ при 4.05~еВ; значне обертання в обидвох сполуках може вказувати на можливе застосування цих матеріалів в оптоелектричних схемах. Досліджені сполуки можуть потенційно використовуватись в оптоелектронних технологіях у спінтронних пристроях.%
	
	\keywords подвійні перовскіти, ефект Керра, оптичні властивості, WIEN2k
	
\end{abstract}

\lastpage	
\end{document}